\documentclass{amsart}

\usepackage{natbib}
\usepackage{caption}
\usepackage{graphicx}

 \usepackage{lineno}
\title{Linking age, survival and transit time distributions}

\begin{document}
\maketitle
	
\begin{abstract}
	Although the concepts of age, survival and transit time have been widely used in many fields, including population dynamics, chemical engineering, and hydrology, a comprehensive mathematical framework is still missing. Here we discuss several relationships among these quantities by starting from the evolution equation for the joint distribution of age and survival, from which the  equations for age and survival time readily follow. It also becomes apparent how the statistical dependence between age and survival is directly related to either the age-dependence of the loss function or the survival-time dependence of the input function. The solution of the joint distribution equation also allows us to obtain the relationships between the age at exit (or death) and the survival time at input (or birth), as well as to stress the symmetries of the various distributions under time reversal. The transit time is then obtained as a sum of the age and survival time, and its properties are discussed along with the general relationships between their mean values. The special case of steady state case is analyzed in detail. Some examples, inspired by hydrologic applications, are presented to illustrate the theory with the specific results.
\end{abstract}

\section{Introduction}

There is considerable interest in estimating age and transit times of elements in a physical system or, equivalently, of individuals in a population, in disciplines as diverse as population dynamics and demography \citep{m1925applications,von1959some,preston2000demography,bongaarts2003estimating}, chemical engineering \citep{nauman1969residence,nauman2008residence}, and hydrology and geophysics (e.g., among many others \cite{eriksson1971compartment,cvetkovic1994transport,goode1996direct,ginn1999distribution,delhez1999toward,mcguire2006review,duffy2010dynamical,botter2011catchment,mcdonnell2014debates,harman2014time,porporato2015probabilistic,benettin2015tracking}). It has in fact become increasingly clear that the age, survival time, and the total time spent by each element in a system may provide additional key insights into specific aspects of a system's behavior. This viewpoint, which can be considered as a time-integrated Lagrangian perspective, has been especially emphasized in groundwater systems \citep{maloszewski1982determining,cvetkovic1994transport,goode1996direct,ginn1999distribution} and in the hydrological response of watersheds \citep{mcguire2006review,mcdonnell2014debates}, using both theoretical \citep{botter2011catchment,velde2012quantifying,harman2014time,benettin2015tracking,porporato2015probabilistic} and field approaches \citep{hrachowitz2009using,birkel2011using,mcdonnell2014debates}.

Recent discussions in the literature about the role of internal variability and external forcing (e.g., rainfall) on the properties of age distributions \citep{porporato2015probabilistic}, as well as the differences between age and survival time distributions, their degree of statistical dependence, and their symmetry under time reversal (e.g., \cite{cornaton2006groundwater,harman2014time,benettin2015tracking}) have made evident that a comprehensive theory of age and related concepts is still missing. Toward this goal, in this contribution we focus on the linkage between age and survival time distribution in both transient and steady-state conditions. Differently from what was assumed in \cite{benettin2015tracking}, we show that age and survival time are in general statistically dependent quantities (the only case of independence being the one of time- and age-independent loss (or input) in steady state). The theoretical framework afforded by the evolution equation of the joint distributions of age and survival also provides a means to easily understand the time symmetries between age and survival, and the derivation of the general properties of the transit-time distribution.

We should warn the readers unfamiliar with the previously cited literature that, perhaps because of the contributions from many disciplines, the terminology which identifies these variables is hardly unified. For example, apart from the age, the definition of which seems uncontroversial ($\tau$ in what follows), the survival time (here indicated as $\sigma$) is often also indicated as life expectancy, while input and output rates are often also called birth and death functions. The variable with possibly 
the most appellations is the so-called transit time ($T$), the sum of age and survival time, which is also indicated as travel time, life span,  total life time, and sojourn time. As long as the mathematical formalism is clear and the notation kept consistent, as we have hopefully done here, we trust that these different names will not confuse the readers.

The paper is organized as follows. The evolution equation for the joint distribution of age and survival time is introduced and solved in section \ref{sec:joint}, with boundary conditions given by the survival time distribution at birth and the related age distribution at death. The solution is used to derive the transit time distribution by a simple integration in section \ref{sec:transit}. The steady state conditions are discussed in section \ref{Sec:steadystate}. Finally, we present some applications in section \ref{sec:applications} with the purpose of showing some interesting details of the theory. While most of these applications have a close connection to hydrological and fluid mechanic systems, they are by necessity highly idealized to allow us to focus on the novel theoretical results, avoiding the additional complications that more realistic applications with random external forcing (e.g, rainfall) and spatial heterogeneities would add.

\section{Joint Distribution of Age and Survival Time}

\label{sec:joint}
The transit time ($T$) of an element of a system is the sum of the time spent since the entrance/birth, called the age ($\tau$), and the time that it will spend before exit/death, called the survival time ($\sigma$). At a given time $t$, each element is characterized by a certain age and survival (and thus transit) time, which globally can be described by the joint distribution $\varphi(t,\tau,\sigma)$. In words, $\varphi(t,\tau,\sigma)d\tau d\sigma$ represents the (infinitesimal) amount of elements (e.g., a mass or population number having age between $\tau$ and $\tau+d\tau$ and survival time $\sigma$ and $\sigma+d\sigma$ at time $t$.

\begin{figure}
	\noindent\includegraphics[width=\columnwidth]{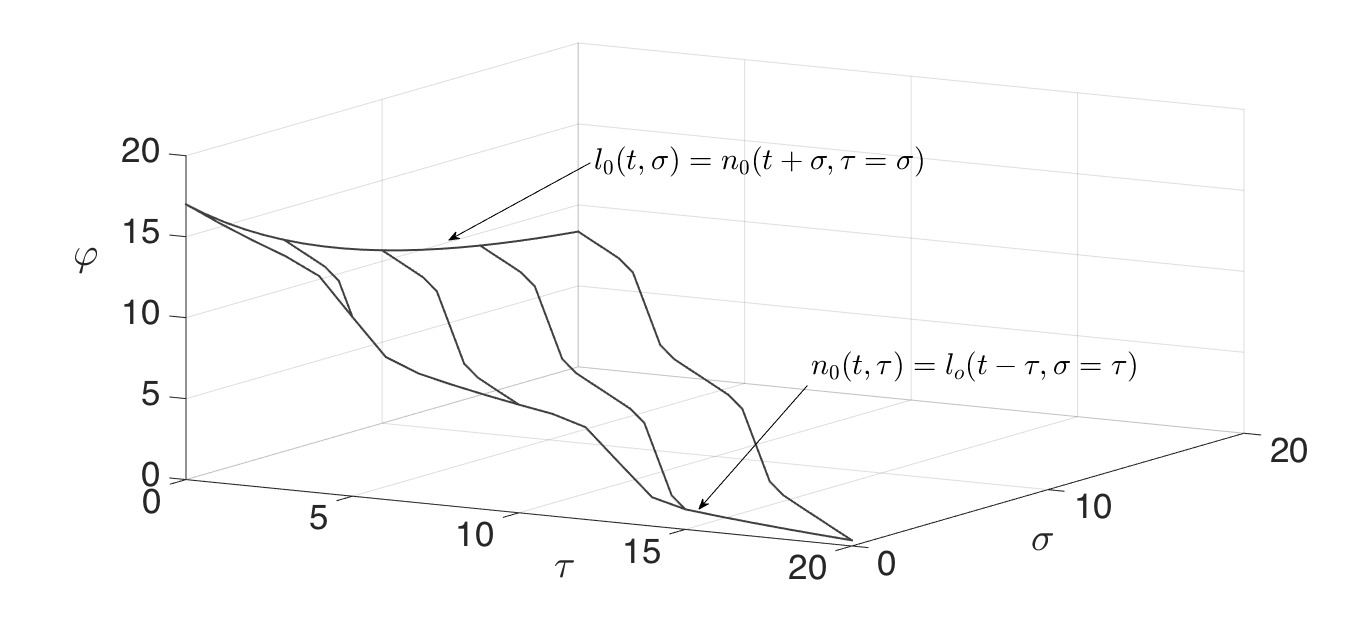}
	\centering
	\caption{\large Example of the joint distribution of age and survival time in transient conditions in which the input has been increased in time, showing how $\varphi$ results from a simple translation of the boundary conditions.}
	\label{Fig:translationBCs}
\end{figure}

The balance equation for the joint distribution $\varphi(t,\tau,\sigma)$ can be obtained considering that, as the system evolves in time, $\varphi(t,\tau,\sigma)$ is conserved along the lines orthogonal to the bisector in the $\tau, \sigma$ plane, which are characterized by having constant $T$.  Based on these considerations, one can readily write
\begin{linenomath*}
	\begin{equation}
	\label{eq:jointeq}
	\frac{\partial \varphi}{\partial t}+\frac{\partial \varphi}{\partial \tau}-\frac{\partial \varphi}{\partial \sigma}=0.
	\end{equation}
\end{linenomath*}
The equation is controlled by the boundary conditions $\varphi(t,\tau,\sigma=0)=l_{0}(t,\sigma)$, which is the survival time distribution at input/birth, and $\varphi(t,\tau=0,\sigma)=n_0(t,\tau)$, which is the age distribution at output/death.
An example of the evolution of the joint distribution is shown in Figure \ref{Fig:translationBCs}, showing how $\varphi$ is simply the input boundary condition on the $\sigma$ axis, translating in time along lines of constant $T$ until it crosses the $\tau$ axis where $\sigma=0$. The figure also clearly shows how the two boundary conditions can not be independent, as will be seen more precisely later. It is interesting that the contribution of input and output to the system is entirely felt through the boundary conditions. In more general cases, elements could also enter with age different from zero (immigration) or exit with a non-zero survival time (emigration), in which case equation (\ref{eq:jointeq}) should also contain corresponding source and sink terms; these generalization however will not be pursued in this paper.

More formally, moving along the characteristic curves, defined by $\frac{d\tau}{ds}=1$, $\frac{d\sigma}{ds}=-1$ and $\frac{dt}{ds}=1$, which are obviously also lines of constant transit time, it is possible to re-express equation (\ref{eq:jointeq}) as
\begin{linenomath*}
	\begin{equation}
	\frac{d \varphi}{d s}=0,
	\end{equation}
\end{linenomath*}
so that the solution is then
\begin{linenomath*}
	\begin{equation}
	\label{eq:jointsol}
	\varphi(t,\tau,\sigma)=\varphi(t-\tau,\tau=0,\sigma+\tau)=\varphi(t+\sigma,\tau+\sigma,\sigma=0),
	\end{equation}
\end{linenomath*}
or equivalently,
\begin{linenomath*}
	\begin{equation}
	\label{eq:jointsol2}
	\varphi(t,\tau,\sigma)=l_{0}(t-\tau,\sigma+\tau)=n_{0}(t+\sigma,\tau+\sigma).
	\end{equation}
\end{linenomath*}
From this it is evident how the joint distribution at a given time is simply the time shift of the boundary conditions and that, if time is reversed, the whole  process is flipped, with the age playing the role of the survival time and vice versa. This type of time symmetry will appear frequently in the following.

\subsection{Age and Survival Time Distributions}

By integrating equation (\ref{eq:jointeq}) over $\sigma$, one obtains the M'Kendrick-von Foerster (MKVF) equation \citep{m1925applications,von1959some,murray2002mathematical,porporato2015probabilistic}, describing the dynamics of an age-structured population equation,
\begin{linenomath*}
	\begin{equation}
	\label{eq.MKVF}
	\frac{\partial n(t,\tau)}{\partial t}+\frac{\partial  n(t,\tau)}{\partial \tau}=-n_{0}(t,\tau),
	\end{equation}
\end{linenomath*}
where $n(t,\tau)=\int_0^\infty \varphi(t,\tau,\sigma)d\sigma$ is the age distribution (mass over time (age)), quantifying the amount of substance having age $\tau$ at time $t$. The sink term, $n_{0}(t,\tau)$, is the age distribution at output/death, previously introduced as a boundary condition for (\ref{eq:jointeq}). It can be written as
\begin{linenomath*}
	\begin{equation}
	\label{eq:defLoss}
	n_{0}(t,\tau)=\mu(t,\tau)n(t,\tau),
	\end{equation}
\end{linenomath*}
where $\mu(t,\tau)$ is the age and mass specific output rate. With initial condition $n(0,\tau)$ and boundary condition $n(t,0)=\iota(t)$, where $\iota(t)$ is the input/birth rate, the solution of equation (\ref{eq.MKVF}) is \citep{trucco1965mathematical,gurtin1974non,keyfitz1997mckendrick}
\begin{linenomath*}
	\begin{equation}
	\label{eq:MKVFsol}
	n(t,\tau)=\left\{\begin{array}{lll}
	n(0,\tau-t) \ e^{-\int_0^t \mu(u,\tau-t+u)du}& & t<\tau\\
	& & \\
	\iota(t-\tau)\ e^{-\int_0^\tau \mu(t-\tau+u,u)du} & & t>\tau.\end{array}\right.
	\end{equation}
\end{linenomath*}
On the other hand, by integrating (\ref{eq:jointeq}) over $\tau$, a corresponding equation for the survival time distribution is obtained,
\begin{linenomath*}
	\begin{equation}
	\label{Eq.MKVFsurv}
	\frac{\partial l(t,\sigma)}{\partial t} - \frac{\partial l(t,\sigma)}{\partial \sigma}= l_{0}(t,\sigma),
	\end{equation}
\end{linenomath*}
where $l(t,\sigma)=\int_0^\infty \varphi(t,\tau,\sigma)d\tau$ quantifies the amount of substance having survival $\sigma$ at time $t$.
The source term $l_{0}(t,\sigma)$ is the survival time distribution at input/birth, which can be expressed as
\begin{linenomath*}
	\begin{equation}
	\label{eq:defBirth}
	l_{0}(t,\sigma)=\beta(t,\sigma)l(t,\sigma),
	\end{equation}
\end{linenomath*}
with $\beta(t,\sigma)$ being the survival time and mass specific birth rate. The boundary condition is the overall output $l(t,\sigma=0)=o(t)$ and the initial condition is $l(0,\sigma)$. As for the MKVF equation, the solution is obtained with the method of characteristics as
\begin{linenomath*}
	\begin{equation}
	\label{eq:MKVFsurvsol}
	l(t,\sigma)=\left\{\begin{array}{lll}
	l(0,t+\sigma) \ e^{\int_0^t \beta(u,t+\sigma-u)du}& & t<\sigma\\
	& & \\
	o(t+\sigma) \ e^{-\int_{0}^{\sigma} \beta(t+\sigma-u,u)du} & & t>\sigma.\end{array}\right..
	\end{equation}
\end{linenomath*}

By integrating again either equation (\ref{eq.MKVF}) over $\tau$ or equation (\ref{Eq.MKVFsurv}) over $\sigma$, the familiar form of the balance equation is obtained,
\begin{linenomath*}
	\begin{equation}
	\label{eq:balance}
	\frac{d w(t)}{d t}=\iota(t)-o(t),
	\end{equation}
\end{linenomath*}
with 
\begin{linenomath*}
	\begin{equation}
	w(t)=\int\limits_{0}^{\infty}n(t,\tau)d\tau=\int\limits_{0}^{\infty}l(t,\sigma)d\sigma
	\end{equation}
\end{linenomath*}
and where the input and output can also be written as
$\iota(t)=\int_0^\infty \beta(\sigma)l(t,\sigma)d\sigma$ and $o(t)=\int_0^\infty \mu(\tau)n(t,\tau)d\tau$.

It should be noted that, when the solution of (\ref{eq:jointeq}) is available, the age and survival distributions, $n(t,\tau)$ and $l(t,\sigma)$, and the evolution of $w(t)$ can be directly obtained by integrating the joint distribution $\varphi(t,\tau,\sigma)$, without need to go through the corresponding equations (\ref{eq.MKVF}), (\ref{Eq.MKVFsurv}), and (\ref{eq:balance}). Looking at these equations, it is also worth noting again the symmetry of the problem with respect to the time reversal, upon which age and survival time exchange their roles, with the output becoming the input and the age-specific loss function playing the part of the survival-specific birth function and vice versa. It is easy to see, in fact, that with these substitutions and $t'=-t$, equation (\ref{eq.MKVF}) and (\ref{Eq.MKVFsurv}) are interchangeable.

\subsection{Statistical Dependence of Age and Survival}

It is possible at this point to establish a relationship between the age-specific output and the survival-specific input and, in turn, discuss the conditional distributions between age and survival times. The latter will be essential in deriving the residence time statistics in section \ref{Sec:steadystate}.
To this purpose, we begin by returning to equation (\ref{eq:jointsol}) which immediately furnishes the relationship between the boundary conditions by setting either $\tau=0$ or $\sigma=0$ (see Figure \ref{Fig:translationBCs}),
\begin{linenomath*}
	\begin{equation}
	\label{eq:relBC}
	\begin{array}{ll}
	l_{0}(t,\sigma)=n_{0}(t+\sigma,\tau=\sigma),  \\
	n_{0}(t,\tau)=l_{0}(t-\tau,\sigma=\tau).
	\end{array}
	\end{equation}
\end{linenomath*}
These equations, when expressed in terms of their probability density functions (PDFs) normalized to have area one,
\begin{linenomath*}
	\begin{equation}
	f_{\tau_0}(t,\tau)=\frac{n_0(t,\tau)}{o(t)} \ \ \ {\rm and} \ \ \ f_{\sigma_0}(t,\sigma)=\frac{l_0(t,\sigma)}{\iota(t)},
	\end{equation}
\end{linenomath*}
become a relationship already obtained by \cite{niemi1977residence},
\begin{linenomath*}
	\begin{equation}
	\iota(t)f_{\sigma_{0}}(t,\sigma)=o(t+\sigma)f_{\tau_0}(t+\sigma,\tau=\sigma).
	\end{equation}
\end{linenomath*}
The subscripts $\tau_{0}$ and $\sigma_{0}$, in particular, refer to the variables age at death and survival time at birth, respectively. They will be used explicitly when it is necessary to refer to them as random variables to distinguish them from age $\tau$ and survival time $\sigma$ of the entire population, as in section \ref{sec:meanvalues}.

From (\ref{eq:relBC}), a relationship between the birth and loss function is then obtained from the definition of $n_{0}$ and $l_{0}$, in equations (\ref{eq:defLoss}) and (\ref{eq:defBirth}), in which the respective solutions for $n(t,\tau)$ and $l(t,\sigma)$ can be substituted from (\ref{eq:MKVFsol}) and (\ref{eq:MKVFsurvsol}), giving
\begin{linenomath*}
	\begin{equation}
	\begin{array}{lll}
	\mu(t,\tau)n(\tau-t) \ e^{-\int_0^t \mu(u,\tau-t+u)du}= \beta(t-\tau,\tau)l(t) \ e^{\int_{0}^{t-\tau} \beta(u,t-u)du}& & t<\tau=\sigma\\
	& & \\
	\mu(t,\tau)\iota(t-\tau)\ e^{-\int_0^\tau \mu(t-\tau+u,u)du}= \beta(t-\tau,\tau) o(t) \ e^{-\int_{0}^{\tau} \beta(t-u,u)du} & & t>\tau=\sigma.
	\end{array}
	\end{equation}
\end{linenomath*}
This clearly shows that the age- and survival-specific birth and loss functions are not independent.

Coming back to equation (\ref{eq:jointsol2}), the joint distribution can be expressed in terms of the conditional distribution between age and survival time. To this purpose, it is more useful to consider the PDFs, instead of distributions, and define
\begin{linenomath*}
	\begin{equation}
	\label{eq:conditional}
	f_{\tau,\sigma}(t,\tau,\sigma)=\frac{\varphi(t,\tau,\sigma)}{w(t)}=f_{\sigma|\tau}(t,\sigma|\tau)\frac{n(t,\tau)}{w(t)}=f_{\tau|\sigma}(t,\tau|\sigma)\frac{l(t,\sigma)}{w(t)},
	\end{equation}
\end{linenomath*}
where $\frac{n(t,\tau)}{w(t)}=f_\tau(t,\tau)$ and  $\frac{l(t,\sigma)}{w(t)}=f_\sigma(t,\sigma)$ are the marginal PDFs of age and survival time, respectively.

Focusing, as an example, on the conditional PDF of survival given age, an expression can be derived combining (\ref{eq:conditional}) and (\ref{eq:jointsol2}),
\begin{linenomath*}
	\begin{equation}
	\label{eq:conddistr}
	f_{\sigma|\tau}(t,\sigma|\tau)=\frac{n_{0}(t+\sigma,\tau+\sigma)}{n(t,\tau)}.
	\end{equation}
\end{linenomath*}
Substituting in (\ref{eq:conddistr}) the solutions for $n_{0}$ and $n$ given by ($\ref{eq:MKVFsol}$) and (\ref{eq:defLoss}), one readily obtains
\begin{linenomath*}
	\begin{equation}
	\label{eq:condsol}
	f_{\sigma|\tau}(t,\sigma|\tau)=\left\{\begin{array}{lll}
	\mu(t+\sigma,\tau+\sigma) e^{-\int_{t}^{t+\sigma}\mu(u,\tau-t+u)du} && t<\tau\\
	& & \\
	\mu(t+\sigma,\tau+\sigma)e^{-\int_{\tau}^{\tau+\sigma}\mu(t-\tau+u,u) du}  & & t>\tau \end{array}\right. .
	\end{equation}
\end{linenomath*}
Only when this expression is equal to the marginal distribution of age, are the age and survival time statistically independent. Thus comparing with the marginal
\begin{linenomath*}
	\begin{equation}
	f_\sigma(t,\sigma)=\left\{\begin{array}{lll}
	\frac{l(0,t+\sigma)}{w(t)} \ e^{\int_0^t \beta(u,t+\sigma-u)du}&& t<\sigma\\
	& & \\
	\frac{o(t+\sigma)}{w(t)} \ e^{-\int_{0}^{\sigma} \beta(t+\sigma-u,u)du} & & t>\sigma \end{array}\right. ,
	\end{equation}
\end{linenomath*}
one sees that, in general, marginal and conditional probability distributions are different. This implies that (when elements are sampled at random in the system) age and survival are typically statistically dependent variables.
This remains true, in transient conditions, even when the loss and birth functions are constant, because the two distributions remain different for $t<\tau$ and $t<\sigma$. Further considerations on (\ref{eq:condsol}) will be given, for steady state conditions in section \ref{Sec:steadystate}.

\section{Transit Time}
\label{sec:transit}

\begin{figure}
	\noindent\includegraphics[width=\textwidth]{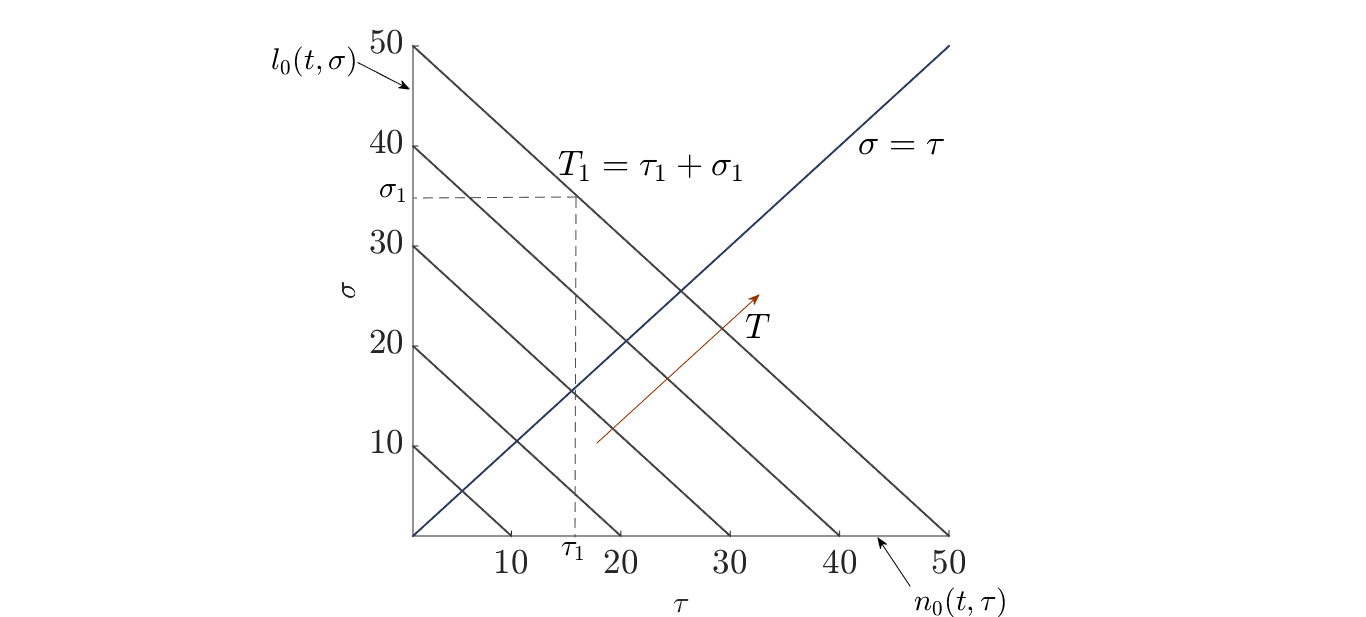}
	\centering
	\caption{ \large Lines at constant transit time shown in the plane $\tau, \sigma$. The boundary conditions $n_{0}(t,\tau_{0})$ and $l_{0}(t,\sigma_{0})$ are also indicated on the $\tau$ and $\sigma$ axes, respectively.}
	\label{Fig:LinesTconst}
\end{figure}

As already said, the transit time, $T$, is the total time spent by an element in the system, given by the sum of age and survival time,
\begin{linenomath*}
	\begin{equation}
	\label{eq:deftransit}
	T=\tau+\sigma.
	\end{equation}
\end{linenomath*}
Its distribution can thus be obtained as the distribution of the sum of the two random variables, the age and survival time, e.g. \cite{springer1979algebra,van2007stochastic},
\begin{linenomath*}
	\begin{equation}
	\label{eq:intjoint1}
	\phi(t,T)=\int_{T<\tau+\sigma<T+dT} \varphi(t,\tau,\sigma) \,d\tau \,d\sigma,
	\end{equation}
\end{linenomath*}
which can also be written as
\begin{linenomath*}
	\begin{equation}
	\label{eq:intjoint2}
	\phi(t,T)=\int\limits_{0}^{T}\varphi(t,\tau',T-\tau')d\tau'=\int\limits_{0}^{T}\varphi(t,T-\sigma',\sigma')d\sigma'.
	\end{equation}
\end{linenomath*}
As shown in Figure \ref{Fig:LinesTconst}, equation (\ref{eq:intjoint2}) is the integral along lines of constant $T$, which are orthogonal to the bisector on the plane $(\tau,\sigma)$. Using equation (\ref{eq:jointsol})  one also has
\begin{linenomath*}
	\begin{equation}
	\label{eq:transdistr}
	\phi(t,T)=\int\limits_{0}^{T} \varphi(t+\sigma',T,0) d \sigma'= \int\limits_{0}^{T} n_{0}(t+\sigma',T) d \sigma'=\int\limits_{0}^{T} \mu(t+\sigma',T) n(t+\sigma',T) d \sigma',
	\end{equation}
\end{linenomath*}
which tells us that we can know the transit time distribution at time $t$ by summing up the amount leaving the system with age $\tau=T$ within the time window $(t,t+T)$. Only when $\tau$ and $\sigma$ are statistically independent is equation (\ref{eq:transdistr}) a simple convolution integral. An alternative expression can be similarly obtained from equation (\ref{eq:jointsol}),
\begin{linenomath*}
	\begin{equation}
	\label{eq:transdistr2}
	\phi(t,T)=\int_{-T}^0 \varphi(t-T+\sigma,0,T)d\sigma'= \int\limits_{-T}^{0} \beta(t+\sigma',T) l(t+\sigma',T) d \sigma',
	\end{equation}
\end{linenomath*}
which instead looks back to the time window ($t-T$,$t$) and sums all the elements entering with survival time $\sigma=T$.

It may be useful to note that the transit time distribution refers in general to any element of the control volume (or population). If one instead only focuses on the elements entering the system (or the newborns), their transit time $T$ is equal to their survival time $\sigma$, because for them $\tau=0$, and their distribution is $l_0(t,\sigma)$. Analogously, focusing on the elements leaving the system (or dying), $T$ equals the age $\tau$ and their residence time distribution is the equal to $n_0(t,\tau)$.

\section{Steady State}
\label{Sec:steadystate}
\subsection{Distributions}
Several of the previous relationships assume an interesting, simplified form at steady state, a necessary condition for which is that $n_{0}$ and $l_{0}$ are time independent. In such a case, the balance equation for a steady state system is simply
\begin{linenomath*}
	\begin{equation}
	\frac{d w}{d t}=0=\iota-o,
	\end{equation}
\end{linenomath*}
and from equation (\ref{eq:jointsol}) it follows that
\begin{linenomath*}
	\begin{equation}
	\label{eq:equalcond}
	n_{0}(\tau)=l_{0}(\sigma=\tau),
	\end{equation}
\end{linenomath*}
meaning that not only the overall input equals the overall output, but also that an input with a fixed age $\tau$ must be balanced by an output of equal survival time $\sigma$. As a result, the joint distribution $\varphi(\tau,\sigma)$ is constant, and equal to $n_{0}(\tau=T)=l_{0}(\sigma=T)$ along lines of constant $T$.

In steady state conditions, age and survival time distributions are derived by taking $t\to\infty$ in (\ref{eq:MKVFsol}) and (\ref{eq:MKVFsurvsol}), that is
\begin{linenomath*}
	\begin{equation}
	\label{eq:agesteady}
	n(\tau)=\iota e^{-\int\limits_{0}^{\tau}\mu(u)du},
	\end{equation}
\end{linenomath*}
and
\begin{linenomath*}
	\begin{equation}
	\label{eq:survsteady}
	l(\sigma)=o e^{-\int\limits_{0}^{\sigma}\beta(u)du}.
	\end{equation}
\end{linenomath*}
In particular, from (\ref{eq:jointsol}), the integral of $\varphi(\tau,\sigma)$ over $\tau$ is the same as the integral over $\sigma$, suggesting that
\begin{linenomath*}
	\begin{equation}
	\label{eq:equaldistr}
	n(\tau)=l(\sigma=\tau).
	\end{equation}
\end{linenomath*}
In turn, it follows immediately that, at steady state, birth and loss functions must be equal,
\begin{linenomath*}
	\begin{equation}
	\label{eq:equallossbirth}
	\mu(\tau)=\beta(\sigma=\tau).
	\end{equation}
\end{linenomath*}

We note that the so-called survivor function \citep{von1959some,cox1962renewal,daly2006state}, defined as the exceedance probability of survival at steady state, can be obtained by dividing either the age distribution or the survival distribution by the input $\iota$ (or by the output $o$),
\begin{linenomath*}
	\begin{equation}
	S(\tau)=\frac{n(\tau)}{\iota}=\frac{l(\sigma=\tau)}{o}=e^{-\int\limits_{0}^{\tau}\mu(u)du}.
	\end{equation}
\end{linenomath*}
so that, as well known \citep{cox1962renewal,trucco1965mathematical},
\begin{linenomath*}
	\begin{equation}
	\label{eq:agedeathsteadynorm}
	-\frac{dS(\tau)}{d\tau}=f_{\tau_{0}}(\tau)=f_{\sigma_{0}}(\sigma=\tau)=\mu(\tau) e^{-\int\limits_{0}^{\tau}\mu(u)du}.
	\end{equation}
\end{linenomath*}

The transit-time distribution at steady state can also be easily obtained by solving the integral in equation (\ref{eq:transdistr}) and substituting equation (\ref{eq:agesteady}),
\begin{linenomath*}
	\begin{equation}
	\label{eq:transdistrsteady}
	\phi(T)=T \mu(T)n(T)=\iota T \mu(T) e^{-\int\limits_{0}^{\tau}\mu(u)du},
	\end{equation}
\end{linenomath*}
or, normalized as a PDF,
\begin{linenomath*}
	\begin{equation}
	\label{eq:transdistrsteadynorm}
	f_{T}(T)=\frac{\iota}{w} T \mu(T) e^{-\int\limits_{0}^{T}\mu(u)du}.
	\end{equation}
\end{linenomath*}

Finally, regarding the conditional probabilities, from equation (\ref{eq:conddistr}),
\begin{linenomath*}
	\begin{equation}
	\label{eq:survconddistrsteady}
	f_{\sigma|\tau}(\sigma|\tau)=\frac{n_{0}(\tau+\sigma)}{n(\tau)}=\mu(\tau+\sigma)e^{-\int\limits_{\tau}^{\tau+\sigma}\mu(u)du}
	\end{equation}
\end{linenomath*}
and
\begin{linenomath*}
	\begin{equation}
	\label{eq:ageconddistrsteady}
	f_{\tau|\sigma}(\tau|\sigma)=\frac{l_{0}(\sigma+\tau)}{l(\sigma)}=\beta(\sigma+\tau)e^{-\int\limits_{\sigma}^{\sigma+\tau}\beta(u)du}.
	\end{equation}
\end{linenomath*}
Because of (\ref{eq:equallossbirth}), the two are obviously equal. 

Comparing for example the distribution (\ref{eq:ageconddistrsteady}) with the corresponding marginal PDF
\begin{linenomath*}
	\begin{equation}
	\label{eq:agesteadynorm}
	f_{\tau}(\tau)=f_{\sigma}(\sigma=\tau)=\frac{\iota}{w}e^{-\int\limits_{0}^{\tau}\mu(u)du},
	\end{equation}
\end{linenomath*}
it becomes clear that only for constant $\mu=\beta$, are (\ref{eq:survconddistrsteady}) and (\ref{eq:ageconddistrsteady}) equal to their marginal distributions, thereby implying that age and survival time are statistically independent. This is essentially due to the rescaling (or memoryless) property of the resulting exponential distributions \citep{ross2014introduction}, which is the form taken by all these distributions in this special case (see section 5.1). In general, however, when the input and loss functions depend respectively on age and survival time, the two variables are statistically dependent, as will be shown in detail in the applications.

\subsection{Mean values}
\label{sec:meanvalues}
In steady state, because of (\ref{eq:equaldistr}), the age distribution and survival distribution have same mean
\begin{linenomath*}
	\begin{equation}
	\bar{\tau}=\bar{\sigma}.
	\end{equation}
\end{linenomath*}
The mean age at death and mean survival time at birth are also equal,
\begin{linenomath*}
	\begin{equation}
	\bar{\tau}_{0}=\bar{\sigma}_{0},
	\end{equation}
\end{linenomath*}
while the mean transit time is then
\begin{linenomath*}
	\begin{equation}
	\label{eq:ageand transit}
	\bar{T}=2 \bar{\tau}=2 \bar{\sigma}.
	\end{equation}
\end{linenomath*}

With regard to the mean transit time, by definition,
\begin{linenomath*}
	\begin{equation}
	\bar{T}=\frac{\int\limits_{o}^{\infty}T \phi(T)dT}{w}=\frac{\int\limits_{o}^{\infty}T^{2}n_{0}(T)dT}{w}=
	\frac{1}{\bar{\tau_{0}}}\frac{\int\limits_{o}^{\infty}T^{2}n_{0}(T)dT}{o},
	\end{equation}
\end{linenomath*}
where the last equality has been obtained by multiplying and dividing by the output $o$.
Now, remembering that
\begin{linenomath*}
	\begin{equation}
	\frac{1}{o}\int\limits_{o}^{\infty}T^{2}n_{0}(T)dT=\frac{1}{o}\int\limits_{o}^{\infty}Tn_{0}(T)dT+\rm{var}(\tau_{0})=
	\bar{\tau_{0}}^{2}+\rm{var}(\tau_{0}),
	\end{equation}
\end{linenomath*}
where $\rm{var(\cdot)}$ is the variance of the respective variable, then one obtains the exact relationship
\begin{linenomath*}
	\begin{equation}
	\label{eq:reltransdeath}
	\bar{T}=\bar{\tau_{0}}+\frac{\rm{var}(\tau_{0})}{\bar{\tau_{0}}}.
	\end{equation}
\end{linenomath*}
Thus, in general, $\bar{T}\geq\bar{\tau_{0}}=\bar{\sigma_{0}}$, so that $\bar{T}$ represents an upper bound for both mean age and mean survival time. In particular, $\bar{\tau_{0}}$ equals $\bar{T}$ only when the loss function is a Dirac delta function, for which the variance of $T$ is zero, as in the case of a plug-flow system.

In addition, substituting expression (\ref{eq:ageand transit}) into the equation above, an exact link between $\bar{\tau}$ and $\bar{\tau_{0}}$ is also obtained as
\begin{linenomath*}
	\begin{equation}
	\label{eq:relageedeath}
	\bar{\tau}=\frac{\bar{\tau_{0}}}{2}+\frac{\rm{var}(\tau_{0})}{2 \bar{\tau_{0}}}.
	\end{equation}
\end{linenomath*}
The same condition was obtained in a somewhat different way by \cite{bjorkstrom1978note}.
Only in the case of $\mu$ and $\beta$ constant, then $\bar{\tau}=\bar{\tau}_{0}$ and $\bar{\sigma}=\bar{\sigma}_{0}$, and
$\bar{T}=2 \bar{\tau}_{0}=2 \bar{\sigma}_{0}$, while for a Dirac delta loss function, $\bar{T}=\bar{\tau_{0}}=\frac{\bar{\tau}}{2}$.

The manner in which input and loss functions depend on age and survival time plays a key role in determining whether the mean of the age and survival time, $\bar{\tau}$ and $\bar{\sigma}$, are greater or lower than the mean age at death and the mean survival time at birth, $\bar{\tau}_{0}$ and $\bar{\sigma}_{0}$. For example, as already discussed by \cite{bolin1973note} and \cite{porporato2015probabilistic}, in the case of a loss function $\mu$ which selects preferably young elements leaving older element to age in the system, the resulting mean age at death is lower than the mean age in the system, i.e., $\bar{\tau}_{0}\textless\bar{\tau}$. On the contrary, the case $\bar{\tau}_{0}\textgreater\bar{\tau}$
is true whenever older elements tend to be chosen by $\mu$, leaving young ones to keep the mean age low in the system, compared to the mean age at death (see section \ref{app4}). 

\section{Applications}
\label{sec:applications}

We present four examples to illustrate the previously discussed theory. The first is a simple steady state system with constant birth and loss functions. The second consists of a plug-flow system in which all the elements have the same transit time. The third application is characterized by a periodicity of the age-independent loss function, while the fourth one focuses on the role of age-dependence in the loss function.

\subsection{Linear System at Steady State}
\label{Sec:linesteady}
\begin{figure}
	\noindent\includegraphics[width=\textwidth]{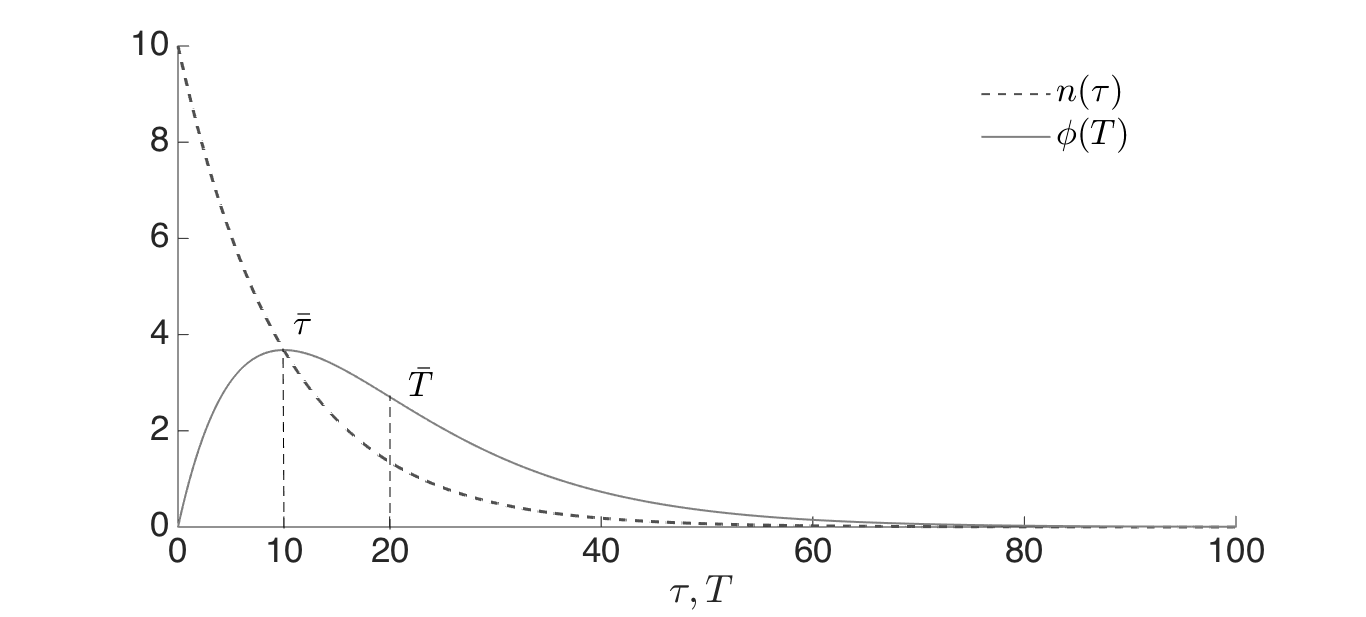}
	\centering
	\caption{ \large Age and transit time distributions for a steady state system with constant and age independent loss function, as discussed in section \ref{Sec:linesteady}. The distributions are calculated for $\iota=10$ and $\eta=0.1$.}
	\label{Fig:Linstead}
\end{figure}

This simplest case, which represents a well-mixed system, serves as a point of reference for the more complex cases presented later. It is characterized by constant and equal birth and loss function,
\begin{linenomath*}
	\begin{equation}
	\mu=\beta=\eta,
	\end{equation}
\end{linenomath*}
so that the balance equation gives
\begin{linenomath*}
	\begin{equation}
	w=\frac{\iota}{\eta},
	\end{equation}
\end{linenomath*}
while the normalized age and survival time distributions are
\begin{linenomath*}
	\begin{equation}
	f_{\tau}(\tau)=\eta e^{-\eta \tau} \ \ \ {\rm and} \ \ \ f_{\sigma}(\sigma)=\eta e^{-\eta \sigma}
	\end{equation}
\end{linenomath*}
with mean $\bar{\tau}=\bar{\sigma}=\frac{1}{\eta}$ (see Figure \ref{Fig:Linstead}). It is easy to show that the same exponential function results from (\ref{eq:survconddistrsteady}) and (\ref{eq:ageconddistrsteady}) for the conditional PDFs, so that in these special case age and survival are statistical independent and their joint distribution is simply the product of the two distributions.

The age distribution at death (survival time distribution at birth) is simply obtained by multiplying $n(\tau)$ by $\eta$ ($l(\sigma)$ by $\eta$),
\begin{linenomath*}
	\begin{equation}
	n_{0}(\tau)=\eta \iota e^{-\eta \tau},
	\end{equation}
\end{linenomath*}
which in its normalized form is equal to the age distribution, $f_{\tau_{0}}(\tau)=\eta e^{-\eta \tau}$, and similarly with $l_0$ and $f_{\tau_{0}}$. The mean values are	
\begin{linenomath*}
	\begin{equation}
	\bar{\tau_{0}}=\bar{\sigma_{0}}=\frac{\theta}{o}=\frac{1}{\eta}.
	\end{equation}
\end{linenomath*}

The transit time distribution (Fig. \ref{Fig:Linstead}) is calculated from equation (\ref{eq:transdistrsteady}),
\begin{linenomath*}
	\begin{equation}
	\phi(T)=T\eta n(T)=T \eta \iota e^{-\eta \tau},
	\end{equation}
\end{linenomath*}
which, written as a PDF, becomes
\begin{linenomath*}
	\begin{equation}
	f_{T}(T)=T \eta^{2} e^{-\eta \tau},
	\end{equation}
\end{linenomath*}
an Erlang-2 distribution for the sum of two independent, exponentially distributed random variables (see Figure \ref{Fig:Linstead}), with mean $\bar{T}=2/\eta$ \citep{cox1962renewal,forbes2011statistical}.

\subsection{Dirac Delta as Survival Time Distribution at Birth}
\label{sec:appdelta}
While in the previous example age and survival time were statistically independent, this application takes the opposite extreme of complete (i.e., deterministic) dependence. To this purpose, we assume that the survival distribution at birth is a Dirac delta on a specific survival time $\sigma^{*}$, modulated sinusoidally in time,
\begin{linenomath*}
	\begin{equation}
	\label{eq:impsurv}
	l_{0}(t,\sigma)= A(t) \delta(\sigma-\sigma^{*}),
	\end{equation}
\end{linenomath*}
with $A(t)=a+b \sin(\omega t)$. Because all the elements exit after the prescribed time $\sigma^{*}$, this example is representative of a time-dependent plug-flow system.

\begin{figure}
	\noindent\includegraphics[width=\textwidth]{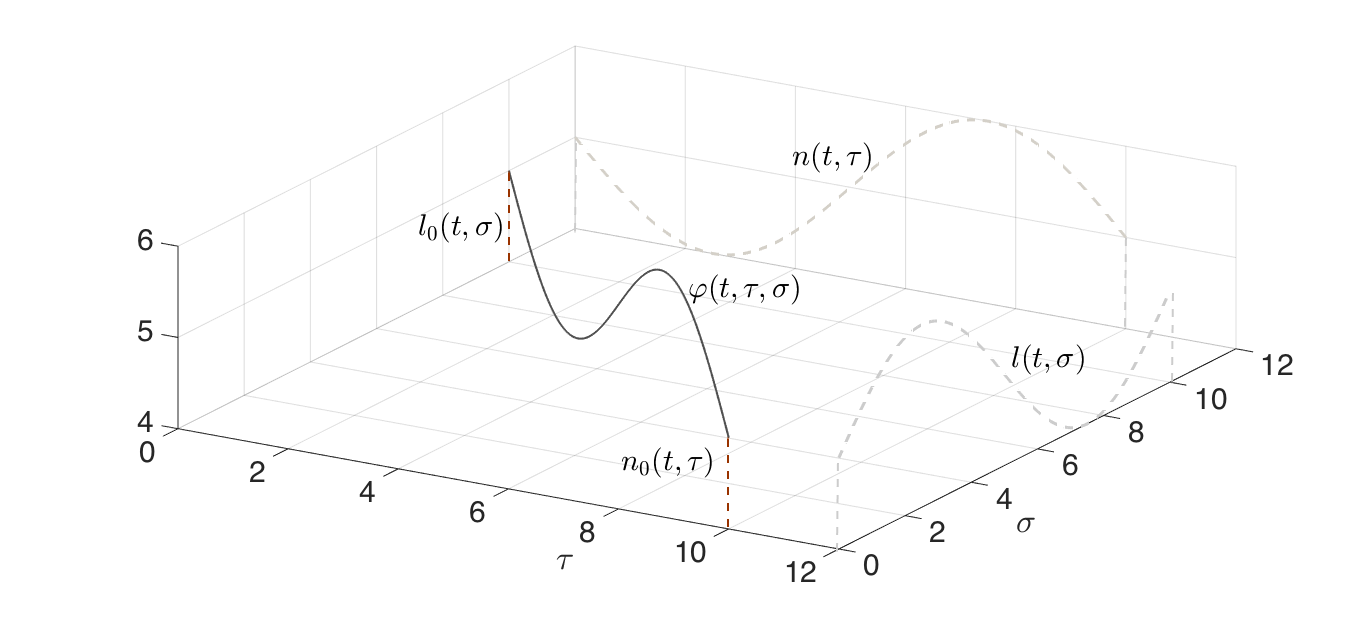}
	\centering
	\caption{ \large Joint distribution of age and survival time for time periodic plug-flow system, with boundary condition $l_{o}(t,\tau)$ imposed as a Dirac delta modulated by the sinusoidal amplitude in (\ref{eq:impsurv}), with $a=5$, $b=1$ and $\sigma^{*}=10$. The marginal distributions of age and survival time are also plotted on the corresponding vertical planes.}
	\label{Fig:deltaapp}
\end{figure}

At the outlet, the age distribution at death $n_{0}$ is the time shifted $l_{0}$, i.e.,
\begin{linenomath*}
	\begin{equation}
	n_{0}(t,\tau) = A(t-\tau) \delta(\tau-\sigma^{*}),
	\end{equation}
\end{linenomath*}
while inside the system, the age distribution is given by
\begin{linenomath*}
	\begin{equation}
	n(t,\tau)=(1-\theta(\tau-\sigma^{*}))A(t-\tau),
	\end{equation}
\end{linenomath*}
where $\theta(\cdot)$ is the Heaviside function \citep{abramowitz2012handbook}. The survival time distribution is
\begin{linenomath*}
	\begin{equation}
	l(t,\sigma)=(1-\theta(\sigma-\sigma^{*}))A(t-\sigma^{*}+\sigma),
	\end{equation}
\end{linenomath*}
while the joint distribution is given by
\begin{linenomath*}
	\begin{equation}
	\varphi(t,\tau,\sigma)=A(t-\tau) \delta(\sigma+\tau-\sigma^{*});
	\end{equation}
\end{linenomath*}
see Figure \ref{Fig:deltaapp}.

The transit time distribution can be calculated from (\ref{eq:transdistr2}),
\begin{linenomath*}
	\begin{equation}
	\phi(t,T)= \int\limits_{-\sigma^{*}}^{0}A(t+x)\delta(T-\sigma^{*})dx=
	\left(a \sigma^{*} +\frac{b (\cos (\sigma^{*}  \omega )-1)}{\omega }\right)\delta(T-\sigma^{*}),
	\end{equation}
\end{linenomath*}
and the conditional distribution,
\begin{linenomath*}
	\begin{equation}
	f_{\sigma|\tau}(t,\sigma|\tau)=f_{\tau|\sigma}(t,\tau|\sigma)=\delta(\sigma+\tau-\sigma^{*}),
	\end{equation}
\end{linenomath*}
which reflects a deterministic relationship between age and survival, imposed by the survival distribution at birth (see Figure \ref{Fig:deltaapp}).

\subsection{Periodic System}

We now consider an extension of the first application, in which the input is constant but the loss function, although still independent of age, is now time periodic,
\begin{linenomath*}
	\begin{equation}
	\label{eq:imposedlossfunction}
	\mu(t)=\eta(t)=a+b \sin(\omega t+\gamma).
	\end{equation}
\end{linenomath*}
From a hydrological point of view, this case is representative of a system with rainfall homogeneously distributed during the year but with seasonally modulated potential evapotranspiration and negligible other losses,
\begin{linenomath*}
	\begin{equation}
	\frac{dw}{dt}=\iota_{0}-\eta(t)w.
	\end{equation}
\end{linenomath*}
The system is still well-mixed, although the transient conditions bring about additional complications that result in statistical dependence between age and survival.

Considering $t \rightarrow \infty$ so that the system has forgotten the initial conditions and has settled on a periodic steady state, the age distribution is from (\ref{eq:MKVFsol})
\begin{linenomath*}
	\begin{equation}
	\label{eq:agedistriperiod}
	n(t,\tau)=\iota_{0}e^{-\int_{0}^{\tau}\eta(t-\tau+u)du}=\iota_{0}e^{-a \tau - \frac{b (\cos (\gamma +\omega  (t-\tau ))-\cos (\gamma +t \omega ))}{\omega }},
	\end{equation}
\end{linenomath*}
while the joint distribution can be obtained through equations (\ref{eq:jointsol2}),
\begin{linenomath*}
	\begin{equation}
	\varphi(t,\tau,\sigma)=\eta(t+\sigma) \iota_{0} e^{-\int_{t-\tau }^{t+\sigma} \eta (u) du}=\eta(t+\sigma) \iota_{0} e^{-a \sigma -a \tau -\frac{b (\cos (\gamma +\omega  (t-\tau ))-\cos (\gamma +\omega  (\sigma +t)))}{\omega }}.
	\end{equation}
\end{linenomath*}

\begin{figure}
	\noindent\includegraphics[width=\textwidth]{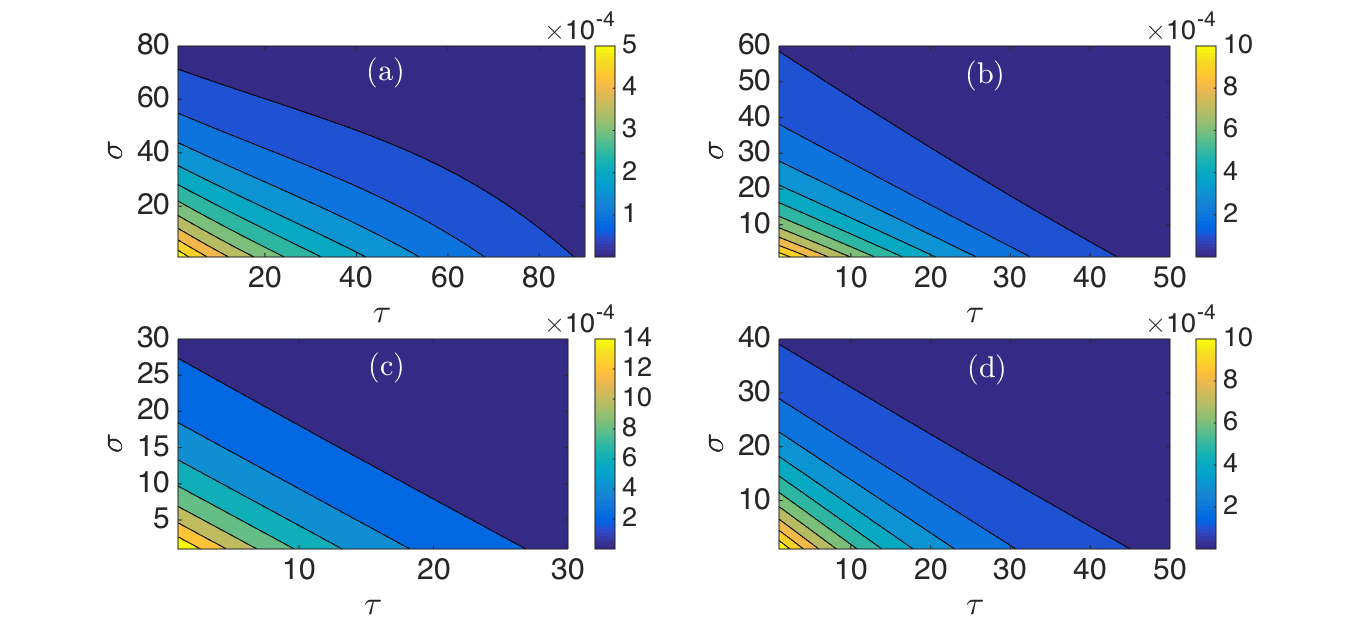}
	\centering
	\caption{\large Joint distributions, for a system with time periodic and age independent loss function (\ref{eq:imposedlossfunction}), calculated at four different times of the year. $t=15$ d (a), $t= 105$ d (b), $t= 196$ (c), $t=288$ d (d). The parameters $a=0.05$, $b=0.025$, $\omega=\frac{2 \pi}{365}$ and $\gamma=-\frac{2 \pi 110}{365}$.}
	\label{Fig:Jointappperiod}
\end{figure}

\begin{figure}
	\noindent\includegraphics[width=\textwidth]{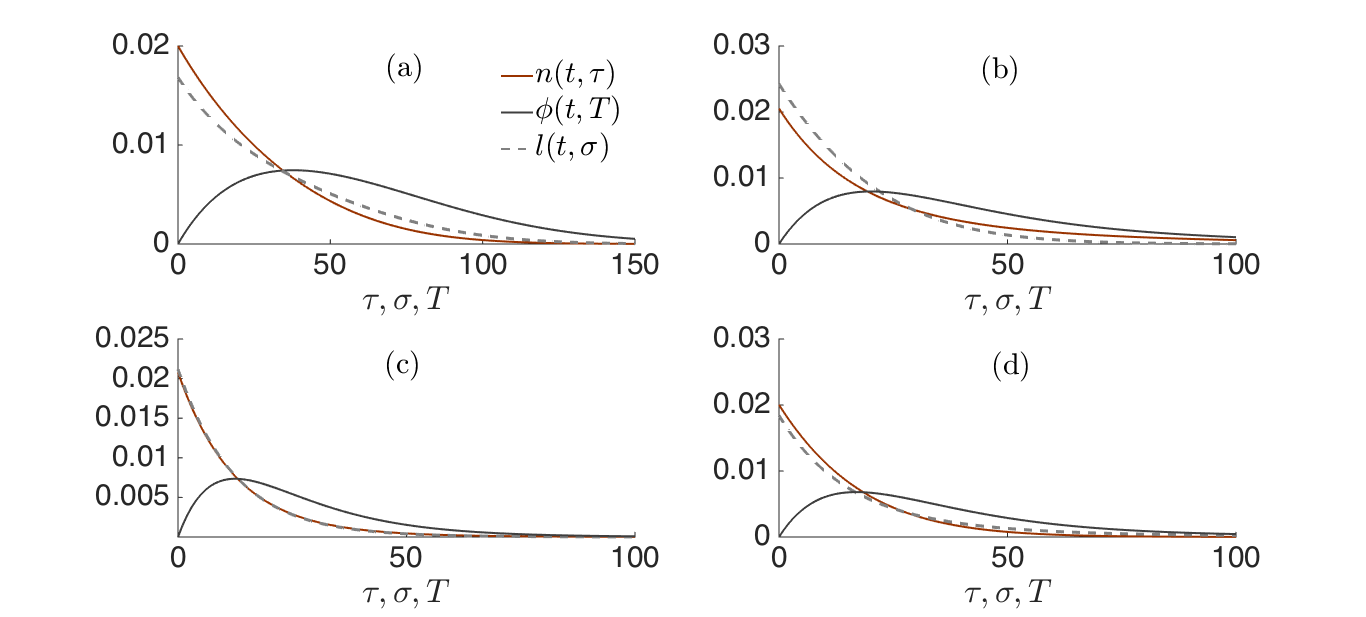}
	\centering
	\caption{ \large Age, survival and transit time distributions for a system with time periodic and age independent loss function (\ref{eq:imposedlossfunction}), calculated at four different times of the year. $t=15$ d (a), $t= 105$ d (b), $t= 196$ (c), $t=288$ d (d). The parameters are the same as in Figure \ref{Fig:Jointappperiod}.}
	\label{Fig:pdfsperiod}
\end{figure}

The joint distribution is plotted in Figure (\ref{Fig:Jointappperiod}) for different days of the year corresponding to the four seasons. They show that the systems has low values of age and survival for the season with high losses (summer), while when the losses diminish (winter), the age and survival start increasing again. Also visible is the asymmetry with respect to the bisector indicating statistical dependence between age and survival induced by the time-varying conditions. The age, survival time and transit time distributions are plotted in Figure \ref{Fig:pdfsperiod}. 

The joint distribution integrated with respect to $\tau$ recovers the survival time distribution $l(t,\sigma)$. However, the integral does not appear to be elementary and here it was only solved numerically.

\begin{figure}
	\noindent\includegraphics[width=\textwidth]{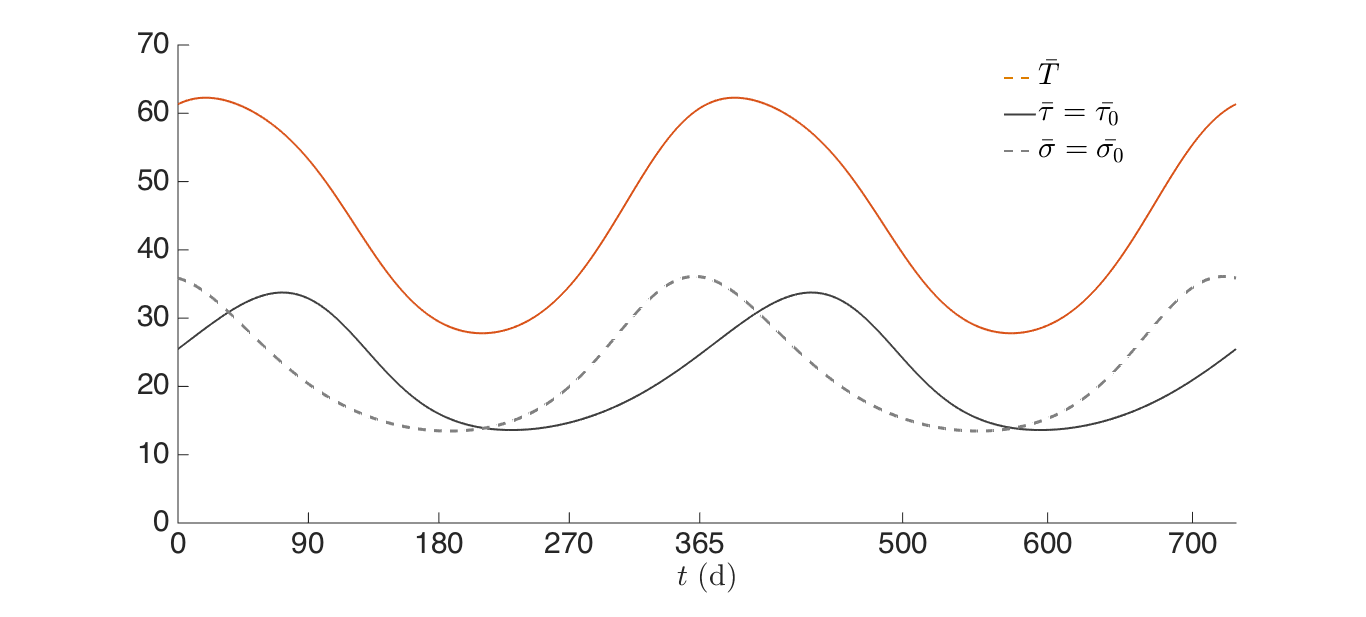}
	\centering
	\caption{ \large Time evolution of the mean age (age at death), mean survival time and survival time at birth and transit time for a system with time periodic and age independent loss function (\ref{eq:imposedlossfunction}).The parameters are the same as in Figure \ref{Fig:Jointappperiod}.}
	\label{Fig:applperiod}
\end{figure}

Finally, regarding the mean values (Figure \ref{Fig:applperiod}), the loss function being age-independent, the mean age at death is equal to the mean age,
\begin{linenomath*}
	\begin{equation}
	\bar{\tau_{0}}=\frac{\eta(t) \int_{0}^{\infty}\tau n(t,\tau) d\tau}{\eta(t) w(t)}=\bar{\tau}.
	\end{equation}
\end{linenomath*}

The mean survival time at birth $\bar{\sigma_{0}}$ was computed analytically, while the mean survival time $\bar{\sigma}$ was computed numerically, and they appeared to be equal.

\subsection{Role of Age-Dependence in the Loss Function}
\label{app4}
In this last example, we analyze the role of the loss function under conditions of steady state. As shown in section \ref{Sec:steadystate}, in steady state, the age and survival time distributions are the same, and therefore similar considerations also apply to a survival-time dependence of the birth function. We consider the age-dependent loss function
\begin{linenomath*}
	\begin{equation}
	\label{eq:imposedlossufunction2}
	\mu(\tau)=\left(\frac{\tau }{c+1}\right)^c,
	\end{equation}
\end{linenomath*}
with $c>-1$. For $-1<c<0$, $\mu$ is a decreasing function, thus selecting younger elements for output, while it is an increasing function of age for $c>0$ with preference for older elements.

The age distributions is obtained from (\ref{eq:agesteady})
\begin{linenomath*}
	\begin{equation}
	n(\tau)= e^{-\left(\frac{\tau }{c+1}\right)^{c+1}}
	\end{equation}
\end{linenomath*}
whereas the age distributions at death is
\begin{linenomath*}
	\begin{equation}
	n_{0}(\tau)=\left(\frac{\tau }{c+1}\right)^c  e^{-\left(\frac{\tau }{c+1}\right)^{c+1}}.
	\end{equation}
\end{linenomath*}
For $-1<c<0$, the age PDF at death is a stretched exponential distribution \citep{sornette2006critical} and, as can be verified through (\ref{eq:relageedeath}), the mean age at death $\bar{\tau_{0}}$ is lower than the mean age $\bar{\tau}$.
When $c>0$, the age PDF at death is a Weibull distribution \citep{kottegoda1997statistics,sornette2006critical} and the mean age at death $\bar{\tau_{0}}$ is greater than the mean age $\bar{\tau}$.
In the limiting case $c=0$, $\mu$ is a constant and the well-mixed system of section \ref{Sec:linesteady} is recovered. In addition, the plug-flow system of section \ref{sec:appdelta} is recovered when taking $c\to\infty$.

\begin{figure}
	\noindent\includegraphics[width=\textwidth]{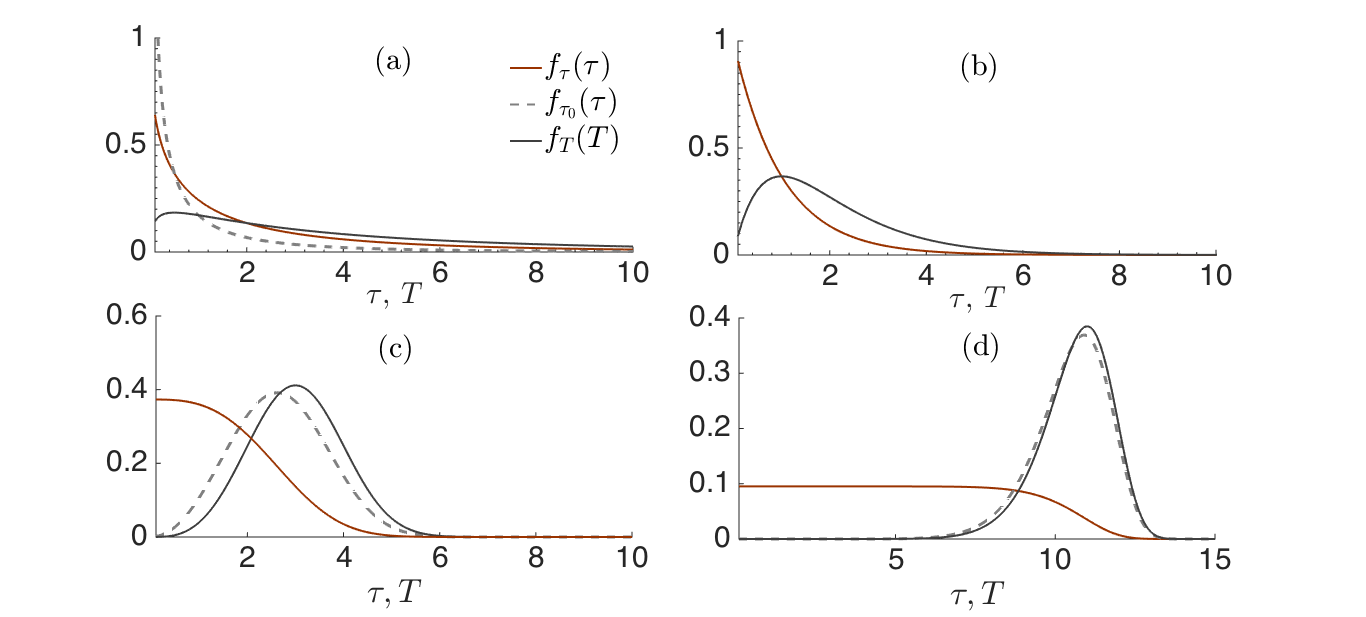}
	\centering
	\caption{ \large Age distribution, age distribution at death and transit time distribution for a steady state system with age dependent loss function given by (\ref{eq:imposedlossufunction2}) with parameter $c=-\frac{1}{2}$ (a), $c=0$ (b), $c=2$ (c) and $c=10$ (d).}
	\label{Fig:distrappl2}
\end{figure}

\begin{figure}
	\noindent\includegraphics[width=\textwidth]{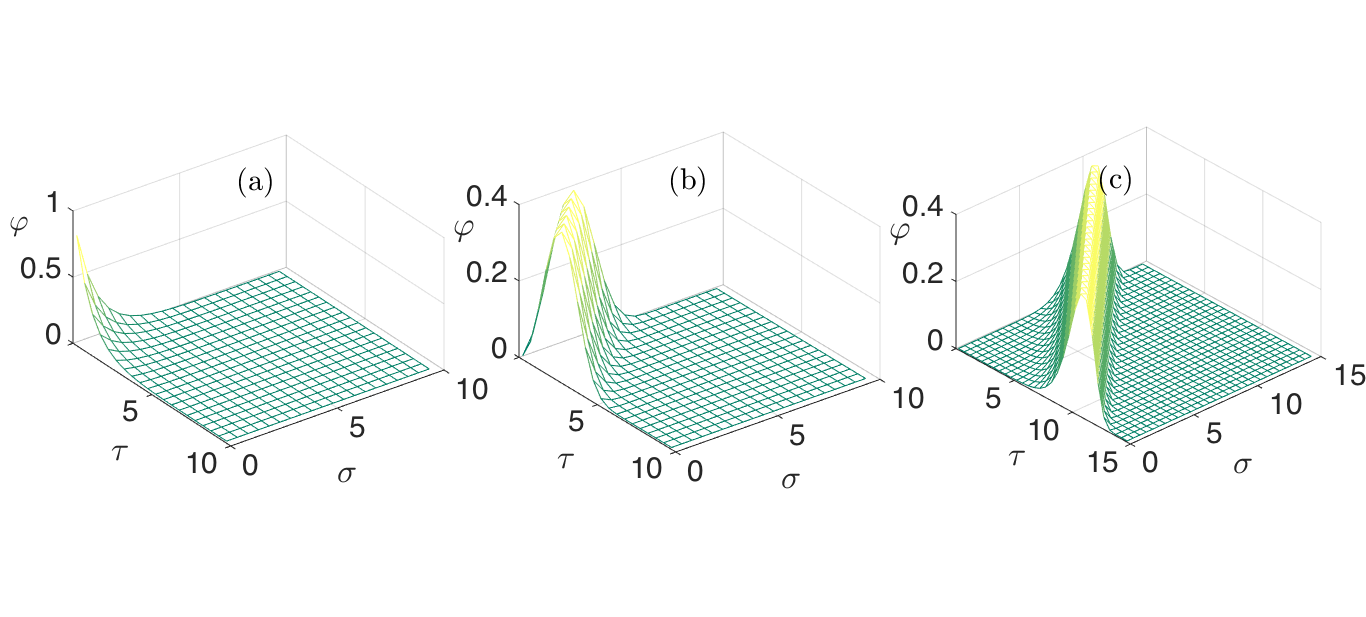}
	\centering
	\caption{\large Joint distribution for a steady state system with age dependent loss function $\mu$ (\ref{eq:imposedlossufunction2}), for $c=0$ (a), $c=2$ (b) and $c=10$ (c).}
	\label{Fig:Applossjoints}
\end{figure}

\begin{figure}
	\noindent\includegraphics[width=\textwidth]{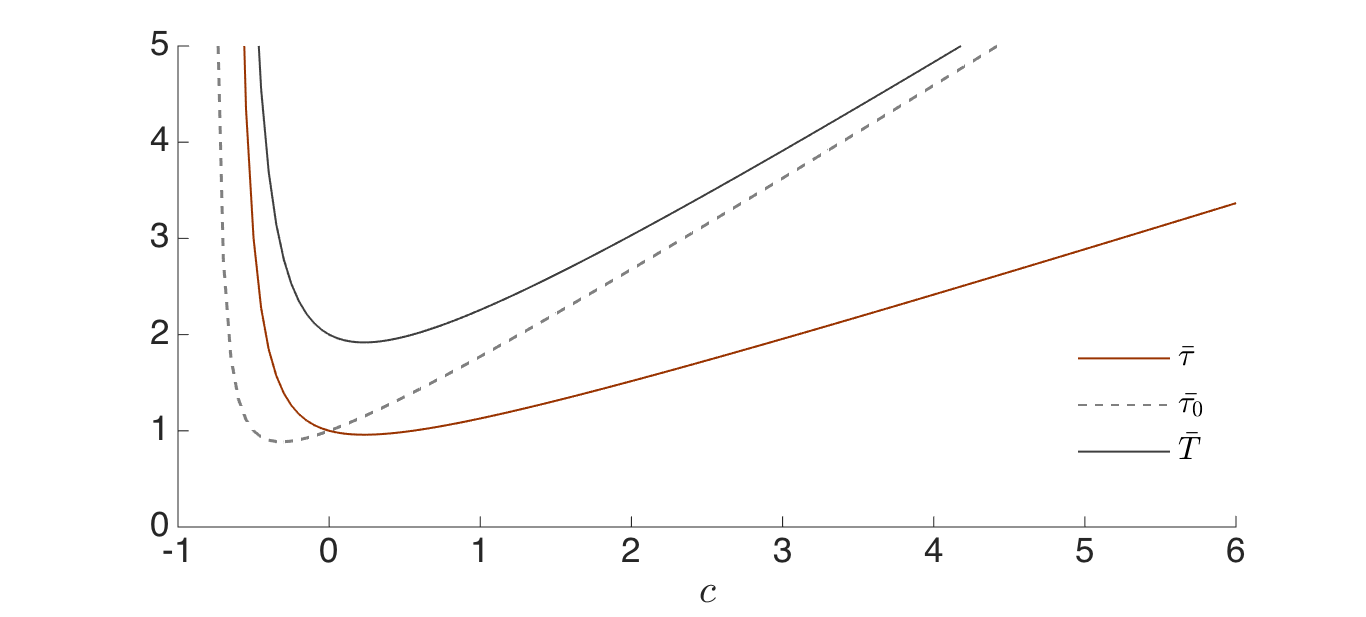}
	\centering
	\caption{ \large Mean transit time, mean age and mean age at death as function of $c$, given the age dependent loss function (\ref{eq:imposedlossufunction2}).}
	\label{Fig:medieofc}
\end{figure}

For different values of $c$, the age, age at death and transit time distributions are plotted in Figure \ref{Fig:distrappl2}, while the joint distributions are plotted in Figure \ref{Fig:Applossjoints}. The mean values are shown in Figure \ref{Fig:medieofc}, where the mean age at death $\bar{\tau_{0}}$ is lower than $\bar{\tau}$ when $-1<c<0$, while it is larger than  $\bar{\tau}$ when $c>0$, and tends asymptotically to $\bar{T}$ for $c\to\infty$. As shown in equation (\ref{eq:reltransdeath}), $\bar{T}$ serves as an upper bound for both mean age and mean age at death.

\section{Conclusions}

Our main results regard the evolution equation of the joint distribution, equation (\ref{eq:jointeq}), which in turn allowed us to obtain the corresponding evolution equation for age (MKVF) and survival time, given by equations (\ref{eq.MKVF}) and (\ref{Eq.MKVFsurv}), respectively. The theory naturally led us to consider the conditional distributions of age and survival, equations (\ref{eq:conditional}), (\ref{eq:conddistr}), and (\ref{eq:condsol}), which helped us clarify some of the statements in the literature about the statistical dependence of these two quantities \citep{cornaton2006groundwater,benettin2015tracking}.
We also obtained general relationships for the transit time distribution, equations (\ref{eq:transdistr}) and (\ref{eq:transdistr2}), and discussed the simplifications induced by the steady state conditions; see equations (\ref{eq:equaldistr}), (\ref{eq:equalcond}), (\ref{eq:equallossbirth}), (\ref{eq:transdistrsteadynorm}) and (\ref{eq:ageconddistrsteady}). Furthermore, we derived exact relationships among the means (\ref{eq:reltransdeath}) and (\ref{eq:relageedeath}), although the latter was already known to \cite{bjorkstrom1978note}.

The present theory is spatially implicit in that it considers, globally, entire populations or finite amounts of substance in a control volume \citep{porporato2015probabilistic}. Interesting future work will consist of connecting it with the spatially explicitly formulation pioneered by  \cite{ginn1999distribution} (and further developed by \cite{weissmann2002dispersion, cornaton2006groundwater,ginn2009notes,cvetkovic2012water}), as done in the case of the MKVF equation by \cite{benettin2013kinematics} and \cite{porporato2015probabilistic}. It also seems promising to explore the use of nonlinear formulations (see \cite{gurtin1974non,gurtin1979some} and related contributions), in which mortality and birth functions also may depend on the total amount $w$, that is $\mu(t,\tau,w)$. Such dependence of birth and mortality on global quantities may be linked to the nonlocal nature of the pressure equation in fluid mechanics \citep{pope2000turbulent,batchelor2000introduction}. Indeed hydrological systems are known to behave such that the loss function at a point is non-local \citep{mcdonnell2014debates}. This is often tacitly assumed in spatially implicit, event-based formulations of rainfall-runoff, which use closure assumptions that depend on the total system storage (\textit{Bartlett et al., 2015}, in preparation). Finally, for realistic hydrologic applications it is necessary to include the effect of stochasticity from the external environment on  the input and output functions \citep{porporato2015probabilistic}. In this case, even the mean quantities become random variables with statistical distributions that may be of great theoretical and practical interest.

\end{document}